\def\begin#1{\csname begin#1\endcsname\ignorespaces}
\def\end#1{\csname end#1\endcsname\ignorespaces}

\font\fourteenrm=cmr10 scaled 1440

\newtoks\slugcomment
\newtoks\lefthead  \let\shorttitle\lefthead
\newtoks\righthead \let\shortauthors\righthead

\def\title#1{\footline={\hfil}
    \line\bgroup\hfil 
    \vbox to 8.3cm
    \bgroup\hsize=9.17cm  
    \baselineskip=12pt\parindent=0pt\parskip=0pt \let\\=\par
    \null\vfil
    {\fourteenrm\baselineskip=18pt
     \textfont0=\fourteenrm \textfont1=\fourteenrm
     #1\bigskip}\bigskip}

\def\author#1{#1\smallskip}
\def\affil#1{#1\vfil}
\def\and{\vfilneg\bigskip}

\def\beginabstract#1\par{\vfil
                         \egroup
                         \hfil\egroup\vfil\noindent}

\def\keywords#1{\noindent {Subject headings: #1}
    \vfil\centerline{\the\slugcomment}\eject\maintextmode}

\def\maintextmode{\headline=
    {\ifodd\pageno\hfil\tenit\the\righthead\quad\hbox{\tenbf\folio}%
     \else\hbox{\tenbf\folio}\quad\tenit\the\lefthead\hfil\fi}}

\parskip \smallskipamount

\def\maybebreak#1{\vskip 0pt plus #1\vsize \penalty -1000
                  \vskip 0pt plus -#1\vsize}

\newcount\secnum
\outer\def\section#1\par{\maybebreak{.1}\bigskip\bigskip\bigskip
      \advance\secnum by 1 \subsecnum=0
      \centerline{\bf\the\secnum. #1}
      \medskip\noindent}
\newcount\subsecnum
\outer\def\subsection#1\par{\maybebreak{.1}\bigskip\bigskip
      \advance\subsecnum by 1
      \centerline{\it\the\secnum.\the\subsecnum. #1}
      \smallskip\noindent}

\def\acknowledgments\par{\bigskip\noindent}

\outer\def\beginreferences{\maybebreak{.1}\bigskip\bigskip\bigskip
      \bgroup\frenchspacing
      \centerline{\bf References}\medskip}
\def\reference#1{\par\hang\noindent\ignorespaces}
\def\endreferences{\par\egroup}

\newcount\itemnum
\def\plainitem#1{\ifhmode\par\fi
                 \noindent\llap{#1\enspace}\ignorespaces}
\def\beginitemize{\ifhmode\par\fi
                  \begingroup \advance\leftskip by \parindent
                  \def\item{\plainitem{$\bullet$}}}
\def\enditemize{\par\endgroup\futurelet\next\par\maybepar}
\def\maybepar{\vskip-\parskip \ifx\next\par\else\noindent\fi}
\def\beginenumerate{\ifhmode\par\fi
                    \begingroup \advance\leftskip by \parindent \itemnum=0
                    \def\item{\advance\itemnum by 1\plainitem{\the\itemnum.}}}

\catcode`\@=11

\newcount\notenum

\def\vfootnote#1{\insert\footins\bgroup\eightpoint
     \interlinepenalty=\interfootnotelinepenalty
     \splittopskip=\ht\strutbox \splitmaxdepth=\dp\strutbox
     \floatingpenalty=20000
     \leftskip=0pt \rightskip=0pt \parskip=1pt \spaceskip=0pt \xspaceskip=0pt
     \smallskip\textindent{#1}\footstrut\futurelet\next\fo@t}

\def\footnote{\global\advance\notenum by 1
    \edef\n@tenum{$^{\the\notenum}$}\let\@sf=\empty
    \ifhmode\edef\@sf{\spacefactor=\the\spacefactor}\/\fi
    \n@tenum\@sf\vfootnote{\n@tenum}}

\catcode`\@=12

\newcount\enum
\def\beginequation{$$\global\advance\enum by 1
                     \gdef\currlabel{\the\enum}}
\def\endequation{\eqno(\the\enum)$$}

\input epsf

\newcount\fignum
\def\beginfigure{\midinsert
                 \global\advance\fignum by 1
                 \def\currlabel{\the\fignum}}
\def\epsscale#1{\epsfxsize=#1\hsize}
\def\plotone#1{\centerline{\epsfbox{#1}}}
\def\caption{\bgroup\noindent{\bf Figure \the\fignum.}
             \aftergroup\par\let\next}

\def\iterate{\let\next=\relax\body\let\next=\iterate\fi\next}

\def\mydef#1#2{\expandafter\ifx\csname#1\endcsname\relax
               \expandafter\def\csname#1\endcsname{#2}\else
               \errmessage{#1 is already \csname#1\endcsname}\fi}

\newif\ifneedrerun
\def\ref#1{\expandafter\ifx\csname#1\endcsname\relax
           ??\needreruntrue \else \csname#1\endcsname\fi}

\def\parb{\par }
\loop
\ifeof1\else
   \ifx\cs\parb\else \expandafter\mydef\cs \fi
\repeat

\immediate\openout2=names

\def\sdef#1#2{\expandafter\xdef\csname#1\endcsname{#2}%
              \immediate\write2{{#1}{#2}}}
\def\label#1{\sdef{#1}{\currlabel}}

\font\eightrm=cmr8
\font\eighti=cmmi8
\skewchar\eighti='177
\font\eightsy=cmsy8
\skewchar\eightsy='60
\font\eightit=cmti8
\font\eightsl=cmsl8
\font\eightbf=cmbx8
\font\eighttt=cmtt8
\def\eightpoint{\textfont0=\eightrm \scriptfont0=\fiverm 
                \def\rm{\fam0\eightrm}\relax
                \textfont1=\eighti \scriptfont1=\fivei 
                \def\mit{\fam1}\def\oldstyle{\fam1\eighti}\relax
                \textfont2=\eightsy \scriptfont2=\fivesy 
                \def\cal{\fam2}\relax
                \textfont3=\tenex \scriptfont3=\tenex 
                \def\it{\fam\itfam\eightit}\relax
                \textfont\itfam=\eightit
                \def\sl{\fam\slfam\eightsl}\relax
                \textfont\slfam=\eightsl
                \def\bf{\fam\bffam\eightbf}\relax
                \textfont\bffam=\eightbf \scriptfont\bffam=\fivebf
                \def\tt{\fam\ttfam\eighttt}\relax
                \textfont\ttfam=\eighttt
                \setbox\strutbox=\hbox{\vrule
                     height7pt depth2pt width0pt}\baselineskip=9pt
                \rm}

\def\frac#1/#2{\mathchoice  {\hbox{$#1\over#2$}} {{#1\over#2}}
               {\scriptstyle{#1\over#2}}
               {#1\mskip-1.5mu/\mskip-1.5mu#2}}
\def\half{{\frac1/2}}
\def\<#1>{\langle#1\rangle}
\def\msd{mass-sheet degeneracy} \def\mdd{mass-disk degeneracy}
\def\vlos{v_{\rm los}}
\def\tgeom{t_{\rm geom}} \def\tgrav{t_{\rm grav}}
\def\Dl{D_{\rm L}} \def\Ds{D_{\rm S}} \def\Dls{D_{\rm LS}}
\def\zl{z_{\rm L}}
\def\tre{\tilde r_{\rm E}} \def\thee{\theta_{\rm E}}
\def\robs{{\bf r}_{\rm obs}}

\font\tenib=cmmib10  \skewchar\tenib='177
\font\sevenib=cmmib7 \skewchar\sevenib='177
\font\fiveib=cmmib5  \skewchar\fiveib='177
\def\bmit{\fam15}
\def\balpha{{\bmit\mathchar"710B}}
\def\bbeta{{\bmit\mathchar"710C}}
\def\btheta{{\bmit\mathchar"7112}}

\slugcomment{To appear in AJ, Oct 2000}

\shortauthors{P. Saha}
\shorttitle{Lensing Degeneracies Revisited}

\begin{document}

\title{Lensing Degeneracies Revisited}

\author{Prasenjit Saha}
\affil{Astronomy Unit, School of Mathematical Sciences\\
Queen Mary and Westfield College\\
London E1~4NS, UK}

\begin{abstract}

This paper shows that the \msd\ and other degeneracies in lensing have
simple geometrical interpretations: they are mostly rescalings of the
arrival-time surface.  Different degeneracies appear in Local Group
lensing and in cosmological lensing, because in the former the
absolute magnification is measured but the image structure is not
resolved, whereas in the latter the reverse usually applies.  The most
dangerous of these is a combination we may call the `\mdd' in
multiply-imaging galaxy lenses, which may lead to large systematic
uncertainties in estimates of cosmological parameters from these
systems.

\end{abstract}


\section{Introduction}

A curious feature of gravitational lensing is that most of the
observables are dimensionless.  This fact leads to some scaleabilities
in lensing theory, which show up as parameter degeneracies when
interpreting observations.  These degeneracies were analyzed in detail
in Gorenstein et al.\ (1988, hereafter G88), elaborating on Falco et
al.\ (1985).  Since that time, while lensing theory has not changed
much the observational situation has changed greatly---recall that in
1988 cluster lensing was still controversial, and Milky Way
microlensing was several years in the future---and with it the
emphasis of theory has shifted.  So it is interesting at this time to
rederive the G88 degeneracies, discuss their current observational
context, and try to gain some new insights into the old results.

G88 discussed three basic degeneracies, which they called the
similarity, prismatic and magnification transformations, and
combinations of them.  The most subtle of these is the
magnification transformation; it seems to have been independently
discovered at least two more times, and is now usually called the
\msd. The present paper will be about the same transformations, but
unlike G88 who started from the lens equation, we will think about
transformations of the arrival-time surface.

\section{The degeneracies}

In fact, lensing degeneracies can all be interpretated as simple
transformations of arrival-time surface
\begin{equation}
t(\btheta) = \tgeom + \tgrav.
\end{equation}
and we can recognize three kinds.
\begin{itemize}
\item `Similarity transformations' scale both $\tgeom$ and
$\tgrav$ by a constant factor. Such operations scale time
delays between images but leave image positions and magnifications
unchanged.
\item The `mass-sheet degeneracy' mixes $\tgeom$ and $\tgrav$ but
in such a way that $t(\btheta)$ is scaled by a constant factor. This
multiplies time delays and all magnifications by a constant factor but
leaves image positions and {\it relative\/} magnifications unchanged.
\item Various other transformations can be written down that
modify $\tgrav$ and possibly also $\tgeom$ in various ways, but leave
$t(\btheta)$ and its derivatives unchanged at all image positions.
These will have no effect on observables, but they cannot be
ignored because they imply uncertainties in what can be inferred about
lenses from the observables.
\end{itemize}
To derive these degeneracies, we start with the full expression
for the arrival time
\begin{equation}
t(\btheta) = \half (1+\zl) {\Dl\Ds\over c\Dls} (\btheta-\bbeta)^2 - 
(1+\zl) {8\pi G\over c^3} \, \nabla^{-2}\Sigma(\btheta),
\label{arriv-eq}
\end{equation}
where $\btheta$ and $\bbeta$ are angular positions on the image and
source planes respectively, $\nabla^{-2}$ denotes the inverse of a
two-dimensional Laplacian, and the other symbols have their usual
meanings.

\subsection{The similarity transformations}

The simplest of the degeneracies appears if the distance factor is
Eq.\ (\ref{arriv-eq}) is unknown (through uncertainty in one or more of
$\Ds$, $\Dl$ or cosmology), which allows the transformation
\begin{equation}
{\Dl\Ds\over\Dls} \to s {\Dl\Ds\over\Dls}, \qquad
\Sigma(\btheta) \to s\Sigma(\btheta).
\label{simd-eq}
\end{equation}
(Here and below $s$ is an arbitrary constant.) G88 call (\ref{simd-eq})
a similarity transformation.  The only effect on observables is to
multiply time delays between images by $s$; neither image positions
nor magnifications change.

If the images are not resolved, another similarity transformation,
\begin{equation}
\btheta \to \sqrt s \btheta, \quad \bbeta \to \sqrt s \bbeta, \qquad
\Sigma(\btheta) \to s\Sigma(\btheta),
\label{sima-eq}
\end{equation}
(not explicitly considered by G88) becomes possible. Here both
sources and images are rescaled by $\surd s$, so magnifications are
unaffected, while once again time delays get multiplied by $s$.  To
avoid a degeneracy of names, I suggest calling Eq.\ (\ref{simd-eq}) a
`distance degeneracy' and Eq.\ (\ref{sima-eq}) an `angular
degeneracy', reserving `similarity transformations' for the whole
category.

The distance and angular degeneracies are independent, in the sense
that it is possible to break one without breaking the other.  Clearly,
one can combine this pair to invent other pairs of independent
similarity transformations.  One such pair, which I suggest calling
the `parallax' and `perspective' degeneracies, are motivated as
follows.

Consider the effect of parallax, i.e., moving the observer.  Say the
observer moves transverse to the optical axis by $\robs$.  For the
observer, the lens will move by $-\robs/\Dl$ and the source by
$-\robs/\Ds$, which amounts to keeping the $\btheta$ fixed and moving
$\bbeta$ by $\robs\Dls/(\Dl\Ds)$.  Applying this change to the arrival
time (\ref{arriv-eq}) and discarding terms with no
$\btheta$-dependence gives
\begin{equation}
t(\btheta) = (1+\zl)\left[
{\Dl\Ds\over c\Dls} \left(\half\btheta^2-\btheta\cdot\bbeta\right)
- {1\over c}\robs\cdot\btheta
- {8\pi G\over c^3} \, \nabla^{-2}\Sigma(\btheta) \right].
\label{arrivr-eq}
\end{equation}
If $\robs$ is known and non-zero, the transformations (\ref{simd-eq})
and (\ref{sima-eq}) are not allowed individually, but the mixture
\begin{equation}
\btheta \to s\btheta, \quad \bbeta \to s\bbeta, \quad
{\Dl\Ds\over\Dls} \to s^{-1} {\Dl\Ds\over\Dls}, \qquad
\Sigma(\btheta) \to s\Sigma(\btheta)
\label{simp-eq}
\end{equation}
is still possible. Here the magnifications will depend on $\robs$, but
Eq.\ (\ref{simp-eq}) does not change them because it rescales
$\btheta$ and $\bbeta$ equally.  I suggest calling Eq.\
(\ref{simp-eq}) the perspective degeneracy because it preserves the
product of the distance and angular scales.  Meanwhile, the similarity
transformation
\begin{equation}
\btheta \to s\btheta, \quad \bbeta \to s\bbeta, \quad
{\Dl\Ds\over\Dls} \to s{\Dl\Ds\over\Dls}, \qquad
\Sigma(\btheta) \to s^3\Sigma(\btheta)
\label{simx-eq}
\end{equation}
is independent of Eq.\ (\ref{simp-eq}) and we can think of it as the
degeneracy that is {\it broken\/} by a parallax observation, so I suggest
calling it the parallax degeneracy.

One usually factors out the similarity transformation by working with
a scaled arrival time surface like so
\begin{equation}
\tau(\btheta) = \half(\btheta-\bbeta)^2 -
                2\nabla_\btheta^{-2}\kappa(\btheta).
\label{tau-eq}
\end{equation}
Here the scaled arrival time $\tau$, the scaled surface density (or
convergence) $\kappa$ and the operator $\nabla_\btheta^{-2}$ are all
dimensionless.  The physical arrival time and density are
\begin{equation}
t(\btheta) = (1+\zl) {\Dl\Ds\over c\Dls} \tau(\btheta), \qquad
\Sigma(\btheta) = {c^2 \over 4\pi G} {\Ds\over\Dls\Dl}\kappa(\btheta).
\label{taudef-eq}
\end{equation}
The usual lensing potential is $\psi=2\nabla_\btheta^{-2}\kappa$ and
the bending angle is $\balpha=\nabla_\btheta\psi$.

\subsection{The mass-sheet degeneracy}

We now rewrite (\ref{tau-eq}) by discarding a $\half\bbeta^2$ term,
since it is constant over the arrival-time surface, and 
using $\nabla_\btheta^2\btheta^2=4$, to get
\begin{equation}
\tau(\btheta) = 2\nabla_\btheta^{-2}(1-\kappa)
              - \btheta\cdot\bbeta.
\label{ntau-eq}
\end{equation}
The transformation
\begin{equation}
1-\kappa \to s (1-\kappa), \qquad \bbeta \to s\bbeta.
\label{msd-eq}
\end{equation}
clearly just rescales time delays while keeping the image structure
the same; but since the source plane is rescaled by $s$ all
magnifications are scaled by $1/s$, leaving relative magnifications
unchanged.  The effect on the lens is to rescale the lensing mass and
then add or subtract a constant mass sheet.  G88 call (\ref{msd-eq}) a
magnification transformation, but `\msd' is its usual
name nowadays.

For a circular lens, the \msd\ preserves the total mass inside an
Einstein radius $\thee$.  We can see this by invoking the
two-dimensional analog of Gauss's flux law in electrostatics, which in
lens notation becomes
\begin{equation}
\oint\balpha\times d{\bf l} = 2\int\!\kappa\,d^2\btheta,
\label{gfl-eq}
\end{equation}
or that the normal component of $\balpha$, integrated along any closed
loop, is proportional to the enclosed mass.  Along an Einstein ring,
$\balpha$ is always radial and hence normal to the ring; also, its
magnitude always equals $\thee$ (since a source at the centre is
imaged onto the ring).  Hence, the left hand integral in Eq.\
(\ref{gfl-eq}) depends only on $\thee$.  Meanwhile the right hand
integral gives twice the enclosed mass.  Thus, fixing the Einstein
radius fixes the enclosed mass.

The \msd\ is broken if there are sources at more than one
redshift. The reason is that we can no longer factor out the
source-redshift dependence as we did in Eqs.\ (\ref{tau-eq}) and
(\ref{taudef-eq}).  Instead, we can replace (\ref{tau-eq}) and
(\ref{taudef-eq}) with 
\begin{equation}
\tau(\btheta) = \half(\btheta-\bbeta)^2 -
                2{\Dls\over\Ds} \nabla_\btheta^{-2}\kappa(\btheta), \qquad
t(\btheta) = (1+\zl) {\Dl\over c} \tau(\btheta), \quad
\Sigma(\btheta) = {c^2 \over 4\pi G} {1\over\Dl}\kappa(\btheta),
\end{equation}
and replace Eq.\ (\ref{ntau-eq}) with
\begin{equation}
\tau(\btheta) = 2\nabla_\btheta^{-2}\left(1-{\Dls\over\Ds}\kappa\right)
              - \btheta\cdot\bbeta,
\label{ntaus-eq}
\end{equation}
Sources at different redshifts imply simultaneous equations of the
type (\ref{ntaus-eq}) but with different factors of $\Dls/\Ds$, which
prevents a transformation like (\ref{msd-eq}).

\subsection{Other degeneracies}

G88 discuss one other transformation, which they call `prismatic',
consisting of adding the same constant to both the source position and
the bending angle.  Physically, this amounts to adding a very massive
lens at very large transverse distance while pushing the source in the
opposite direction.  So it is not as important as the similarity and
magnification transformations.

\begin{figure}
\epsscale{0.6}
\plotone{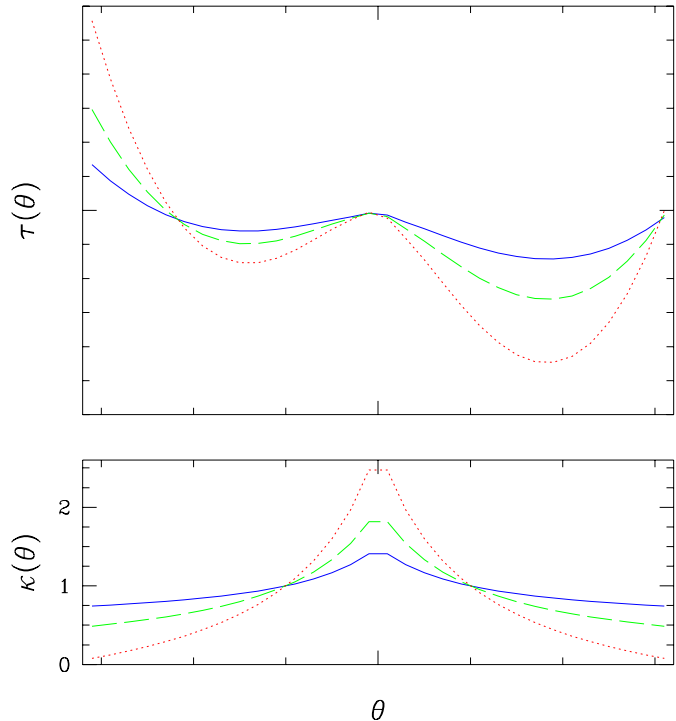}
\caption{Illustration of the \mdd, showing the surface density (lower
panel) and the arrival time (upper panel) for three circular lenses.
The units, except for $\kappa$, are arbitrary.  The arrival time
indicates a saddle point (looking like a local minimum in this cut),
a maximum, and a minimum.  The dashed curves correspond to a
non-singular isothermal lens.  Stretching the time scale amounts to
making lens profile steeper (dotted curves) and shrinking the time
scale amounts to making the lens profile shallower (solid curves).
Note that there is a limit to stretching, because otherwise $\kappa$
will become negative somewhere in the region with images---negative
$\kappa$ {\it outside\/} that region can always be avoided by adding
an external monopole.  But there is no limit to shrinking.}
\label{mdd-fig}
\end{figure}

Clearly, one can concoct any number of localized transformations that
leave $t(\btheta)$ and its derivatives unchanged at all image
positions and do not make $\kappa$ negative anywhere.  An obvious one
is what we may call a `monopole' transformation: any circularly
symmetric redistribution of mass inwards of all observed images, and
any circularly symmetric change in mass outside all observed images
will have no effect on observables.  A more subtle example, which
causes an ambiguity between close and wide binary lenses in Local
Group lensing, is discussed by Dominik (1999).

The monopole transformation has an important indirect effect: it
changes the \msd\ into a `\mdd'---as long as the disk is larger than
the region of images, a circular disk and an infinite sheet are
equivalent in lensing---and a much more dangerous effect, since it
cannot be eliminated by the requirement that $\kappa$ goes to 0 at
large $\btheta$.  Figure \ref{mdd-fig} illustrates.

\section{Digression: velocity dispersions}

Though not a lensing observable, velocity dispersion is often measured
in connection with lensing, and is worth discussing here.

In any lens having approximately critical density, a typical internal
velocity $v$ satisfies
\begin{equation}
v \sim c\thee^\half.
\label{v-eq}
\end{equation}
To derive this, we write $R$ for the lens's size and $M$ for its
mass, and recall that $\thee\sim R/\Dl$ for critical density, $v^2\sim
GM/R$ from the virial theorem, and that $\thee\sim GM/(c^2\Dl)$.

The most familiar example of the scaling (\ref{v-eq}) is for an
isothermal lens.  As is well known, this lens derives from a stellar
dynamical sphere with constant velocity dispersion $\sigma$: the
density is $\rho=\sigma^2/(2\pi Gr^2)$, which amounts to a projected
density of $\Sigma=\sigma^2/(2G\Dl^2\theta^2)$, leading to
\begin{equation}
\thee = 4\pi{\sigma^2\over c^2}{\Dl\over\Ds}.
\label{sigiso-eq}
\end{equation}
One can use Eq.\ (\ref{sigiso-eq}) to define a formal $\sigma$ for any
approximately circular lens.  This formal $\sigma$ can usefully serve
as a surrogate for $\thee$.  Moreover, because of the relation
(\ref{v-eq}) the formal $\sigma$ will be of order the internal
velocities in the lens, but in general it will be different from the
actual velocity dispersion.

To elaborate, let us consider the relation between observed velocity
dispersions and mass distribution.  For a stellar system with no
rotation or other streaming motions, the virial theorem states that
$\<v^2> = \<{\bf r}\cdot\nabla\Phi>$, where $v$ is the stellar
velocity, the averages $\<.\,.>$ are over the stellar
distribution function, and $\Phi$ is the total gravitational
potential.  Thus far there are no symmetry assumptions.  If, however,
spherical symmetry does apply then the line-of-sight direction must
contribute the same as the orthogonal directions, and hence
\begin{equation}
\<\vlos^2> = \frac1/3 \<{\bf r}\cdot\nabla\Phi>,
\label{vlos-eq}
\end{equation}
$\vlos$ being the line-of-sight stellar velocity. If $\Phi$ is due to
an isothermal sphere with dispersion $\sigma$ then ${\bf
r}\cdot\nabla\Phi=2\sigma^2$ everywhere, which gives
\begin{equation}
\<\vlos^2> = \frac2/3\sigma^2
\label{vlosiso-eq}
\end{equation}
(cf.\ equation 4.6 in Kochanek 1993).  For other spherical lenses
one may compare $\<\vlos^2>$ with the formal $\sigma^2$ derived from
Eq.\ (\ref{sigiso-eq}).  For example, consider a homogeneous sphere of
stars of radius $R$, with no non-stellar matter.  Using (\ref{vlos-eq}), we
have
\begin{equation}
\<\vlos^2> = {GM\over 5R}
\label{vloshs-eq}
\end{equation}
where $M$ is the total mass.  If this sphere is barely compact, $M$
and $R$ satisfy $4GM\Dls/(c^2\Dl\Ds)=\thee^2$ and $R/\Dl=\thee$.
Inserting these values into (\ref{vloshs-eq}) and then eliminating
$\thee$ using (\ref{sigiso-eq}) gives
\begin{equation}
\<\vlos^2> = \frac\pi/5\sigma^2,
\label{vlosbchs-eq}
\end{equation}
only 6\% different from (\ref{vlosiso-eq}).

The above relations imply two things: (i)~Eq.\ (\ref{vlos-eq})
indicates that there is a considerable range allowed in $\<\vlos^2>$
for lenses with given formal $\sigma$, and (ii)~Eq.\ (\ref{vlosbchs-eq})
shows that even if $\<\vlos^2>$ is observed to have the expected
isothermal value, it does not follow that the lens is isothermal.  And
all this uncertainty is present without even considering ellipticity
and velocity anisotropy.

In summary, for galaxy and cluster lenses an order-of-magnitude
relation of the type
\begin{equation}
\thee \simeq 2'' \times {\<\vlos^2>\over(300\rm\,km\,s^{-1})^2}
\end{equation}
is useful, but velocity dispersion is not a precise constraint unless
the mass distribution is already known.  Perhaps lenses become much
better constrained if there is much more detailed velocity
information; the answer seems unknown, but see Dejonghe \& Merritt
(1992) and Romanowsky \& Kochanek (1999).

\section{Appearances of the degeneracies in Local Group Lensing}

In Local Group lensing, the similarity transformations are relevant.
The \mdd\ does not apply because absolute magnifications are always
measured, and anyway there are no disk-like lens components involved.

In most Local Group microlensing events only the magnification as a
function of time is measured; the distances are unknown and the image
structure is unresolved, so both distance and angular degeneracies
(alternatively, both parallax and perpective degeneracies) apply.
Note that the events being time-dependent and hence furnishing a whole
sequence of arrival-time surfaces does not prevent the similarity
transformations---for each event one can scale the whole sequence of
arrival-time surfaces by the same factor.

In a few cases ($\sim15$ out of $\sim500$ events observed so far) one
degeneracy has been broken through additional observational
information, and there are prospects for breaking the degeneracies
completely in future with the help of observations from satellites.
The requirements for degeneracy-breaking are well known, but it is
interesting to interpret them in terms of the similarity transformations.

\subsection{Proper motions}

Proper motion measurements break the angular degeneracy
(\ref{sima-eq}), leaving the spatial degeneracy (\ref{simd-eq}).

The significance of proper motion measurements was already appreciated
by Refsdal (1966b), though the configuration then envisaged (lensing
of visible one star by another, with separate proper-motion
measurements for both) is not now considered realistic.  A more
realistic situation, independently pointed out by Gould (1994a),
Nemiroff \& Wickramasinghe (1994), and Witt \& Mao (1994), is of a
lens transiting the source star, in which case the finite size of the
source will flatten the peak of the light curve; if the angular size
of the source star can be estimated then $d\bbeta/dt$, and hence
$\thee$, can be inferred.  Alcock et al.\ (1997) observed such an
event.  Another situation which enables $d\bbeta/dt$ to be measured is
when not the lens itself but a caustic of a binary lens crosses the
source.  Albrow et al.\ (1999, 2000a,b) Afonso et al.\ (2000) and
Alcock et al.\ (2000) have made such measurements.

\subsection{Parallaxes}

For a parallax observation, one needs to introduce the effect of a
suitable $\robs$ in Eq.\ (\ref{arrivr-eq}), and this can be brought
about in two ways.  One way, suggested by Refsdal (1966b) and Gould
(1992, 1994b, 1995), is to have separate observers, using one or more
satellites.  The other way, suggested by Gould (1992), is to exploit
the Earth's acceleration.  Now, a constant $d\robs/dt$ is irrelevant
in Eq.\ (\ref{arrivr-eq}) because it can be absorbed inside
$d\bbeta/dt$. But a known $d^2\robs/dt^2$ modifies both the
magnification (photometric parallax) and the proper motion of the
image centroid (astrometric parallax).  Photometric parallax events
have been observed by Alcock et al.\ (1995), Bennett et al.\ (1997)
and Mao (1999).

Parallax observations leave the perspective degeneracy
(\ref{simp-eq}), which holds the combination
\begin{equation}
{\Dl\Ds\over\Dls} \Sigma(\btheta)
\label{tildere-eq}
\end{equation}
constant.  For a single mass, (\ref{tildere-eq}) is $\propto\tre^2$
($\tre$ being the Einstein radius projected onto the observer plane).
Thus we recover the well-known result that parallax measurements
determine $\tre$.

\subsection{Prospects for combining proper motion and parallax}

As will be clear from the above (and elsewhere---Refsdal 1966b already
made this point) combining proper motion and parallax measurements
will lift all the degeneracies, and enable the lens mass to be solved
for completely.  Prospects for combined measurements have been
discussed in several papers.  Paczy\'nski (1998) and Boden et al.\
(1998) suggest using interferometry to measure both proper motion and
astrometric parallax, while Miyamoto \& Yoshii (1995), Berlinski \&
Saha (1998) and Gould \& Salim (1999) advocate using interferometry
for proper motions and photometry for parallaxes.

\section{Appearances of the Degeneracies in Cosmological Lensing}

The degeneracies in cosmological lensing are complementary to those in
Local Group lensing.  The angular degeneracy does not appear because
there is always some resolved image structure.  The distance
degeneracy appears, but in a very simple way---with redshifts usually
measurable, the distance factor in Eq.\ (\ref{arriv-eq}) is $\propto
H_0^{-1}$ times a weak and readily-quantifiable dependence on
cosmology.  The main thing to worry about is the \mdd.  The following
very briefly discusses the various contexts.

\subsection{In quasar microlensing}

In quasar microlensing the \mdd\ is actually useful!  Modeling
microlensing of lensed quasars involves computing lightcurves in a
potential of stars plus smooth matter.  In such computations (e.g.,
Refsdal \& Stabell 1997) a standard trick uses the \mdd\ to transform
away the effect of the smooth matter by rescaling the stellar masses
appropriately---see Eq.\ (24) of Paczy\'nski (1986), which appears to
be an independent discovery of the degeneracy.

The angular degeneracy is also present, because while one obviously
cannot change the angular scale of the macro-images, it is not
forbidden to rescale the micro-image system within each macro-image,
along with the source's proper motion.

\subsection{In cluster lensing}

Another independent discovery of the the \mdd, this time in the
context of cluster lensing, was by Schneider \& Seitz (1994).
Kaiser's (1995) formula expressing $\nabla\ln(1-\kappa)$ in terms of
observable ellipticities makes the degeneracy particularly explicit:
multiplying $(1-\kappa)$ by a constant will not change the
ellipticities.  These and later papers have drawn considerable
attention to the need to break the degeneracy, and research towards
this end is active.  Most of the effort is directed towards using the
magnification information from number counts of background galaxies
(see e.g., Taylor et al.\ 1998), but AbdelSalam et al.\ (1998) use the
information that comes from having a range of source redshifts.

\subsection{In quasar macrolensing}

Although lensing degeneracies were originally discovered in the
context of quasar macrolensing, recent literature in this area (e.g.,
the article on `Modeling Galaxy Lenses' by Blandford et al., 2000)
usually does not discuss degeneracies.  The reason, perhaps, is that
the popular parametrized lens models have focused attention on their
respective parameters and away from the global transformations that
produce degeneracies.

Considerable work has been done on fitting parametric models to the
detailed image structure (e.g., Kochanek 1995, Kochanek et al., 2000).
Such work often produces precisely constrained values for the radial
density gradient and the core radius.  But---and this is very
important---those values are conditional upon particular parametrized
lens models, because (a)~the \mdd\ allows one to change the radial
density gradient drastically without changing the image structure at
all, and (b)~the monopole degeneracy makes core radii if anything more
free.  Nor, as we saw in the previous section, do velocity dispersion
measurements provide strong independent constraints unless the mass
profile is assumed already known.

Thus, the mass-disk and monopole degeneracies point to some
significant uncertainties in our current knowledge of galaxy lens
profiles, and hence to uncertainties in estimates of cosmological
parameters from quasar lensing.

The effect of the \mdd\ on estimates of $h$ was already fully
appreciated in Falco et al.\ (1985).  Given a lens model that
reproduces all observations of a lensed quasar and its host galaxy,
one is still free to stretch or shrink the scale of the arrival-time
surface (i.e., $h^{-1}$) using the \mdd---see Figure
\ref{mdd-fig}. There is a limit to stretching, because eventually
$\kappa$ somewhere will reach zero; this means that there is an upper
limit on the inferred $h$.  There is no limit to shrinking the
arrival-time surface: the lens can get arbitrarily close to a disk
with $\kappa=1$ and the inferred $h$ will get closer and closer to
zero!  To prevent this happening one must incorporate some assumptions
about the steepness of the mass profile.  Model-builders are familiar
with such behavior (see e.g., Wambsganss \& Paczy\'nski 1994, Williams
\& Saha 2000).

Degeneracies are even more dangerous for inferences of $\Omega$ and
$\Lambda$ from lensing, because individual lenses contain no
information on these parameters, only the ensemble of lenses
does.\footnote{In a little known companion paper to the famous Refsdal
(1964) on time delays and $h$, Refsdal (1966a) suggested that
time-delays for systems at different redshifts could be put on a sort
of Hubble diagram to determine the other cosmological parameters.  But
at present, researchers prefer to fit the redshift-dependencies of the
density of multiple-image systems and the distribution of image
separations, which also depend on cosmology; these two quantities are
much easier to observe than time delays, but more awkward to interpret
because magnification bias enters.} Several researchers (Maoz \& Rix
1993, Kochanek 1996, Park \& Gott 1997, Chiba \& Yoshii 1999) have
attempted to constrain $\Omega$ and $\Lambda$ from the
redshift-dependence of the source density or the image separations, or
both.  The results are conditional upon different assumptions made by
the various authors, and in particular upon very specific lens
profiles.  Williams (1997) studies the dependence of the
image-separation statistics on lens profiles, and concludes that it is
much larger than the dependence on cosmology. For example, by making
the lens profiles less steep in the inner regions she can make
small-separation systems more magnified and hence (because of
magnification bias) more abundant at high redshifts, thus completely
drowning out the effect of cosmology.\footnote{The image-separation
statistics are complicated by the existence of a number of
wide-separation quasar pairs at low redshifts, with no visible lens.
These are currently thought to be binary quasars with no lensing
involved (Kochanek et al.\ 1999), though spectral similarities in some
cases cast doubts upon that interpretation (Small et al.\ 1997, Peng
et al.\ 1999).  Park \& Gott (1997) find that the image-separation
statistics with the wide-separation pairs included as lenses is not
reproducible using power-law lenses.  Williams (1997) finds that the
same statistics can be reproduced if the lensing galaxies have
changing logarithmic density profiles and follow the scaling laws
characteristic of spirals rather than ellipticals, and concludes that
the resolution of the nature of the wide-separation pairs as lenses or
binaries will lead to constraints on the lensing population. Kochanek
et al.\ (1999) say that Williams's examples are ``inconsistent with
the known properties of galaxies and lenses'' but do not explain which
known properties.}

\section{Summary}

Lensing degeneracies can be simply understood as rescalings or other
transformations of the arrival-time surface that leave various image
properties unaffected.  The most important of these are as follows.
\begin{itemize}
\item `Similarity transformations' typically arise in microlensing.
There are two independent ones which, depending on context, are
usefully taken as:
\begin{enumerate}
\item a `distance' degeneracy where the distance scale varies while the
angular scale stays fixed; and
\item an `angular' degeneracy where the angular scale varies while the
distance scale stays fixed;
\end{enumerate}
or as
\begin{enumerate}
\item a `perspective' degeneracy where the product of the distance and angular
scale varies while the ratio stays fixed; and 
\item a `parallax' degeneracy where the ratio stays fixed while the
product varies.
\end{enumerate}
\item The `\mdd' is typical of cosmological lensing.  It rescales
$$ 1 - {\<\hbox{density}>\over\<\hbox{critical density}>} $$
within a finite disk larger than observed region, in the process
rescaling the total magnification and the time delays, but otherwise
leaving images unaffected.
\item `Localized' degeneracies do not change the arrival-time surface
at image positions, but change it elsewhere; no lensing observable is
altered, but other properties such as core radii and velocity
dispersions may be.
\end{itemize}

The most insidious of these is the \mdd\ when it appears in
multiply-imaging galaxy lenses, where it translates into a serious
source of uncertainty in estimates of $h$ from time delays,
and even worse uncertainties in estimates of $\Omega$ and $\Lambda$
from image statistics.

\acknowledgments

I am grateful to the referee for a number of detailed comments and
suggestions.

\end{document}